# SYSTEMATIC COMPARABLE COMPANY ANALYSIS AND COMPUTATION OF COST OF EQUITY USING CLUSTERING

**Mohammed Perves[1]**

[1] Physics and Computer Science, Wilfrid Laurier University
[1] Lazaridis School of Business & Economics, Wilfrid Laurier University

Advisors
**Pooja Vashisth[2], Jiashu (Jessie) Zhao[3]**

[2] Computer Science, University of Toronto
[3] Physics and Computer Science, Wilfrid Laurier University

[1] moha3440@mylaurier.ca

**ABSTRACT**

Computing cost of equity for private corporations and performing comparable company analysis (comps) for both public and private corporations is an integral but tedious and time-consuming task, with important applications spanning the finance world, from valuations to internal planning. Performing comps traditionally often times include high ambiguity and subjectivity, leading to unreliability and inconsistency. In this paper, I will present a systematic and faster approach to compute cost of equity for private corporations and perform comps for both public and private corporations using spectral and agglomerative clustering. This leads to a reduction in the time required to perform comps by orders of magnitude and entire process being more consistent and reliable.

## 1. Introduction

In the world of finance, Comparable Company Analysis (or "Comps" for short) is a valuation methodology that looks at financial ratios of similar public companies, called "peers" and uses them to derive the value of another company, either public or private. Comps is a relative form of valuation, meaning the value of a company of interest is derived with respect to its peers, unlike other valuation methods such as the discounted cash flow (DCF) analysis, which is an intrinsic form of valuation. In practice, Comps is the most widely used valuation method due to the relative ease in performing the procedure and the required data such as financial ratios being readily available, especially for public companies. Investment bankers, sell-side research analysts, private equity investors, and other market analysts all use Comps to value an IPO, assess the attractiveness of a potential investment (from a value perspective), and value private corporations. However, traditional Comps also have their disadvantages. The primary disadvantage being the fact that Comps are fundamentally influenced by temporary market conditions or non-fundamental factors. Other disadvantages include difficulties in finding appropriate comparable companies for various reasons, rendering the analysis useless when there are few or no comparable companies.

Cost of Equity is the rate of return a company pays out to equity investors. Corporations use cost of equity to assess the relative attractiveness of investments, including both internal projects and external acquisition opportunities. Companies typically use a combination of equity and debt financing [6]. A firm's Weighted Average Cost of Capital (WACC) represents its blended cost of capital across all sources, including common shares, preferred shares, and debt. The cost of each type of capital is weighted by its percentage of total capital and they are added together [7]. The WACC formula is below.

***Where:***
E = Market value of the firm's equity (market cap)
D = Market value of the firm's debt
V = Total value of capital (equity plus debt)
Re = Cost of equity (required rate of return)
Rd = Cost of debt (yield to maturity on existing debt)
T = Tax rate of the firm

$$WACC = \frac{E}{E+D} * R_E + \frac{D}{E+D} * R_D * (1-T)$$

The Weighted Average Cost of Capital serves as the discount rate for calculating the Net Present Value (NPV) of a business. It is also used to evaluate investment opportunities, as it is considered to represent the firm's opportunity cost. Thus, it is used as a hurdle rate by companies [7]. A company will commonly use its WACC as a hurdle rate for evaluating mergers and acquisitions (M&A), as well as for financial modeling of internal investments. If an investment opportunity has a lower Internal Rate of Return (IRR) than its WACC, it should buy back its own shares or pay out a dividend instead of investing in the project [7]. Investors also use WACC as the discount rate when performing a DCF valuation of a firm. It is really easy to compute the WACC for public companies. To compute WACC for public companies you can get all the data you need such as the "E, D, V, T, and Rd" from the company's financial statements and Stock Exchange. To compute the remaining "Re = Cost of Equity", you can use the Capital Asset Pricing Model (CAPM) or the Dividend Capitalization Model (for companies that pay out dividends). CAPM considers the riskiness of an investment relative to the market [6]. The CAPM model is below:

***Where:***
Ri = Expected return on asset i
Rf = Risk-free rate of return
βi = Beta of asset i
E(Rm) = Expected market return

$$R_i = R_f + \beta_i * (R_m - R_f)$$

The Dividend Capitalization Model (also known as the Gordan Growth Model) only applies to companies that pay dividends, and it also assumes that the dividends will grow at a constant rate. The model does not account for investment risk to the extent that CAPM does (since CAPM requires beta) [6].

*Where:*
Re = Cost of Equity
D1 = Dividends/share next year
P0 = Current share price
g = Dividend growth rate

$$GGM = \frac{Do(1 + g)}{ke - g}$$

To compute the WACC for private corporations we cannot use these methods directly because private corporations don't have a Beta or pay out dividends. It is rather a tedious task and often times the computed WACC is suboptimal and doesn't reflect its true value because of the challenges around computing the cost of equity. The cost of equity for private corporations is calculated using the Capital Asset Pricing Model. The firm's beta is calculated by taking the firm's industry average beta [8]. This often times leads to suboptimal values because the average of the industry may not reflect the firm's true Beta. Alternatively, the cost of equity for private corporations is also calculated by simply taking the average of the industry. Again, this leads to suboptimal values because the average of the industry may not reflect the firm's cost of equity.

## 2. Performing Comps Traditionally

Below is a detailed procedure to performs Comps traditionally from the Corporate Finance Institute:

### 2.1 Find the right comparable companies

This is the first and probably the hardest (or most subjective) step in performing a ratio analysis of public companies. The very first thing an analyst should do is look up the company you are trying to value on CapIQ or Bloomberg so you can get a detailed description and industry classification of the business.

The next step is to search either of those databases for companies that operate in the same industry and that have similar characteristics. The closer the match, the better.

The analyst will run a screen based on criteria that include:
  1) Industry classification

2) Geography
3) Size (revenue, assets, employees)
4) Growth rate
5) Margins and profitability

**2.2 Gather financial information**

Once you've found the list of companies that you feel are most relevant to the company, you're trying to value it's time to gather their financial information. Once again, you will probably be working with Bloomberg Terminal or Capital IQ and you can easily use either of them to import financial information directly into Excel. The information you need will vary widely by industry and the company's stage in the business lifecycle. For mature businesses, you will look at metrics like EBITDA and EPS, but for earlier stage companies you may look at Gross Profit or Revenue. If you don't have access to an expensive tool like Bloomberg or Capital IQ you can manually gather this information from annual and quarterly reports, but it will be much more time-consuming.

**2.3 Set up the comps table**

In Excel, you now need to create a table that lists all the relevant information about the companies you're going to analyze.

The main information in comparable company analysis includes:
- Company name
- Share price
- Market capitalization
- Net debt
- Enterprise value
- Revenue
- EBITDA
- EPS
- Analyst estimates

The above information can be organized as shown in our example comparable companies analysis shown below.

| Company Name | Market Data | | | Financial Data | | | | Valuation | | | |
| --- | --- | --- | --- | --- | --- | --- | --- | --- | --- | --- | --- |
| | Price ($/share) | Market Cap ($M) | TEV ($M) | Sales ($M) | EBITDA ($M) | EBIT ($M) | Earnings ($M) | EV/Sales x | EV/EBITDA x | EV/EBIT x | P/E x |
| The Coca-Cola Company | 38.14 | 168,041 | 185,122 | 46,854 | 13,104 | 11,127 | 7,381 | | | | |
| Pepsico, Inc. | 81.37 | 123,883 | 143,824 | 66,415 | 12,344 | 9,878 | 5,618 | | | | |
| Dr Pepper Snapple Group, Inc. | 52.31 | 10,326 | 12,764 | 5,997 | 1,319 | 1,103 | 620 | | | | |
| Monster Beverage Corporation | 69.62 | 11,618 | 11,004 | 2,246 | 606 | 584 | 357 | | | | |
| National Beverage Corp. | 20.81 | 964 | 968 | 645 | 78 | 66 | 41 | | | | |

*Source: https://corporatefinanceinstitute.com/resources/knowledge/valuation/comparable-company-analysis*

## 2.4 Calculate the comparable ratios

With a combination of historical financials and analyst estimates populated in the comps table, it's time to start calculating the various ratios that will be used to value the company in question.

The main ratios included in a comparable company analysis are:
- EV/Revenue
- EV/Gross Profit
- EV/EBITDA
- P/E
- P/NAV
- P/B

| Company Name | Market Data | | | Financial Data | | | | Valuation | | | |
|---|---|---|---|---|---|---|---|---|---|---|---|
| | Price ($/share) | Market Cap ($M) | TEV ($M) | Sales ($M) | EBITDA ($M) | EBIT ($M) | Earnings ($M) | EV/Sales x | EV/EBITDA x | EV/EBIT x | P/E x |
| The Coca-Cola Company | 38.14 | 168,041 | 185,122 | 46,854 | 13,104 | 11,127 | 7,381 | 4.0x | 14.1x | 16.6x | 22.8x |
| Pepsico, Inc. | 81.37 | 123,883 | 143,824 | 66,415 | 12,344 | 9,878 | 5,618 | 2.2x | 11.7x | 14.6x | 22.1x |
| Dr Pepper Snapple Group, Inc. | 52.31 | 10,326 | 12,764 | 5,997 | 1,319 | 1,103 | 620 | 2.1x | 9.7x | 11.6x | 16.7x |
| Monster Beverage Corporation | 69.62 | 11,618 | 11,004 | 2,246 | 606 | 584 | 357 | 4.9x | 18.1x | 18.9x | 32.5x |
| National Beverage Corp. | 20.81 | 964 | 968 | 645 | 78 | 66 | 41 | 1.5x | 12.5x | 14.6x | 23.5x |
| Average | | | | | | | | 2.9x | 13.2x | 15.3x | 23.5x |
| Median | | | | | | | | 2.2x | 12.5x | 14.6x | 22.8x |

*Source: https://corporatefinanceinstitute.com/resources/knowledge/valuation/comparable-company-analysis*

### 2.5 Disadvantages of Traditional Comps

There are several disadvantages to performing comps traditionally. First, it is time consuming. Analysts often time spends hours and days finding similar companies via stock screening and reading company descriptions in hopes of establishing a peer group. Second, the peers determined can vary from analyst to analyst. For instance, one Analyst may think Facebook is a peer of Microsoft because they are both tech companies and another may think the opposite. Lastly, the peer group may not be comprehensive. Since, analyst do manual search for peers, often times, they may miss some peers and their resulting peer group may not be the most accurate.

# 3. Clustering

Clustering is the process of grouping or segmenting a set of objects such that objects within a group are more similar to each other than the objects in other groups. For instance, taking a classroom of students and putting them into k groups based on their favorite sports. It is commonly used in statistical data analysis, with many real-world applications extending across many sectors. There are a variety of algorithms designed to perform this process. The algorithms differ in the way a "Cluster" is defined and how the clusters are found. For most algorithms, the common parameters are similarity or distance functions and the number of clusters.

## 3.1 Types of Clustering

Clustering methods can be divided into two basic types: hierarchical and partitional clustering. Hierarchical clustering either merges smaller clusters into larger clusters or splits larger clusters into smaller clusters. This is typically used if the underlying structure behind the data is a tree and is presented in a dendrogram. Partitional clustering, by contrast, directly partitions the data set into a set of disjoint clusters. Below is helpful table of visuals and summary for some of the clustering methods.

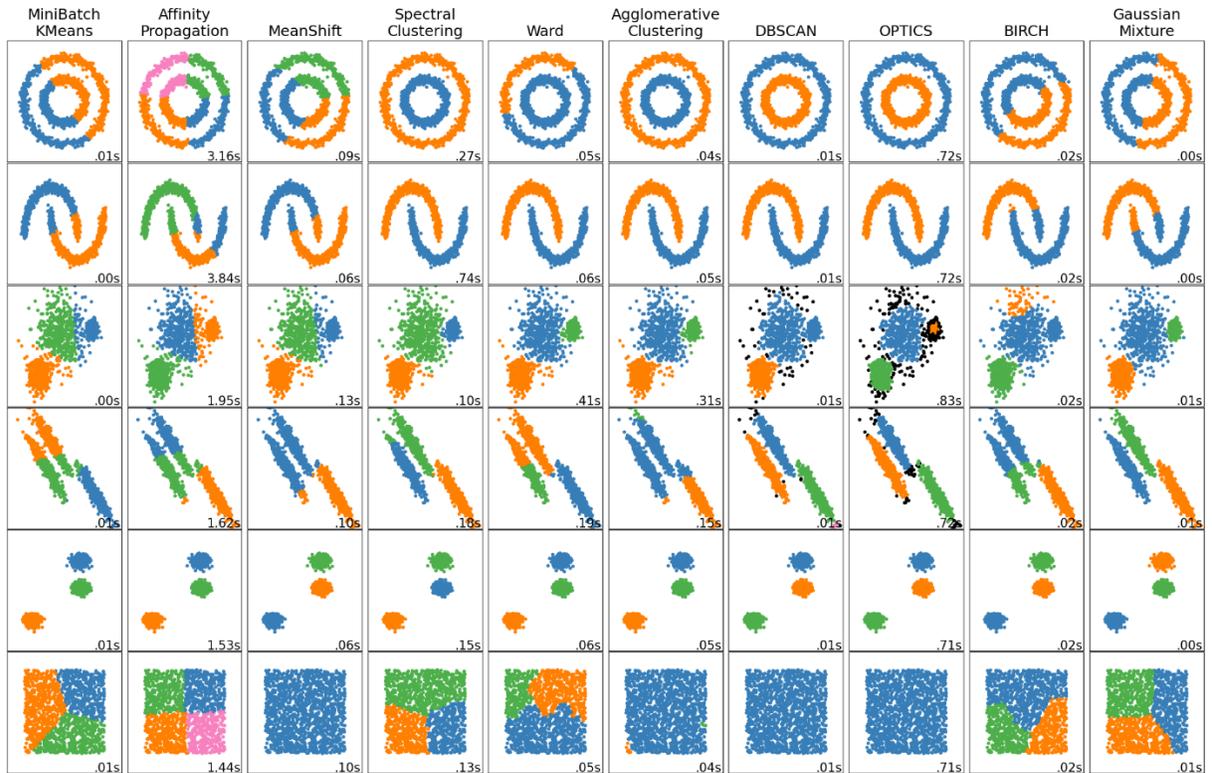

*Source: https://scikit-learn.org/stable/modules/clustering.html*

| Method name | Parameters | Scalability | Usecase | Geometry (metric used) |
|---|---|---|---|---|
| **K-Means** | number of clusters | Very large n_samples, medium n_clusters with MiniBatch code | General-purpose, even cluster size, flat geometry, not too many clusters, inductive | Distances between points |
| **Affinity propagation** | damping, sample preference | Not scalable with n_samples | Many clusters, uneven cluster size, non-flat geometry, inductive | Graph distance (e.g. nearest-neighbor graph) |
| **Mean-shift** | bandwidth | Not scalable with n_samples | Many clusters, uneven cluster size, non-flat geometry, inductive | Distances between points |
| **Spectral clustering** | number of clusters | Medium n_samples, small n_clusters | Few clusters, even cluster size, non-flat geometry, transductive* | Graph distance (e.g. nearest-neighbor graph) |
| **Ward hierarchical clustering** | number of clusters or distance threshold | Large n_samples and n_clusters | Many clusters, possibly connectivity constraints, transductive* | Distances between points |
| **Agglomerative clustering** | number of clusters or distance threshold, linkage type, distance | Large n_samples and n_clusters | Many clusters, possibly connectivity constraints, non Euclidean distances, transductive* | Any pairwise distance |
| **DBSCAN** | neighborhood size | Very large n_samples, medium n_clusters | Non-flat geometry, uneven cluster sizes, outlier removal, transductive* | Distances between nearest points |
| **OPTICS** | minimum cluster membership | Very large n_samples, large n_clusters | Non-flat geometry, uneven cluster sizes, variable cluster density, outlier removal, transductive* | Distances between points |
| **Gaussian mixtures** | many | Not scalable | Flat geometry, good for density estimation, inductive | Mahalanobis distances to centers |
| **BIRCH** | branching factor, threshold, optional global clusterer. | Large n_clusters and n_samples | Large dataset, outlier removal, data reduction, inductive | Euclidean distance between points |

* Transductive clustering methods (in contrast to inductive clustering methods) are not designed to be applied to new, unseen data. *Source: https://scikit-learn.org/stable/modules/clustering.html*

### 3.2 Clustering Algorithms used in the Paper

In this paper, I will be using Spectral and Agglomerative Clustering for several reasons. First, document clustering by nature is hierarchical, with very complex shapes. Agglomerative clustering is great at hierarchical clustering and Spectral clustering is very useful when the structure of the individual clusters is highly non-convex, or more generally when a measure of the center and spread of the cluster is not a suitable description of the complete cluster, such as when clusters are nested circles on the 2D plane [Sklearn].

# 4. Systematic Comps using Clustering

The idea is very simple. Take all the publicly listed stocks (or a very large subset) and obtain as much data as possible, including Wikipedia pages, financial data, such as financial ratios, fundamentals and SEC filings primarily annual 10-K filings. Clean and preprocess the dataset. Then, cluster all the stocks using their Wikipedia pages or SEC filings that outline what the business does (business description), history and so forth; I will call this business description (busdesc) clustering. Now, to perform public comps for a given stock, I simply find which Cluster it belongs to and return the $n^{th}$ nearest neighbors as its peers.

**The proposed approach is as follows:**
1. Create the busdesc dataset by scraping Wikipedia and EDGAR
2. Retrieve the financial dataset from Wharton Research Data Services
3. Perform any required preprocessing and cleaning
    a. Remove non-alpha characters
    b. Perform text tokenization
    c. Remove unwanted words that impedes clustering using regular expressions and manually. For example, removing company names that are nouns, such as "Stanley". There are 2 stocks in the S&P, "Morgan Stanley" and "Stanley Black & Decker Inc" who both have the noun "Stanley" in the Wikipedia pages. If we performed text clustering, these 2 stocks are put in the same clustered because they both contain "Stanley" dozens of times. However, these companies are not similar at all. Morgan Stanley is a multinational investment bank and financial services company, whereas, Stanley Black & Decker Inc is a manufacturer of industrial tools and household hardware and provider of security products.
    d. Remove stop words such as ['a', 'the', 'of', 'for', ...]
    e. Perform stemming
4. Vectorize documents using methods such as Bag of Words, Doc2Vec, Bidirectional Encoder Representations from Transformers (BERT) or Universal Sentence Encoder. I will be using Doc2Vec in combination with TF-IDF in this paper.
5. Perform document clustering using K-Means and Spectral Clustering
6. Find n peers using k-nearest neighbors (KNN) algorithm for any given stock
7. Perform Public Comps using the peers

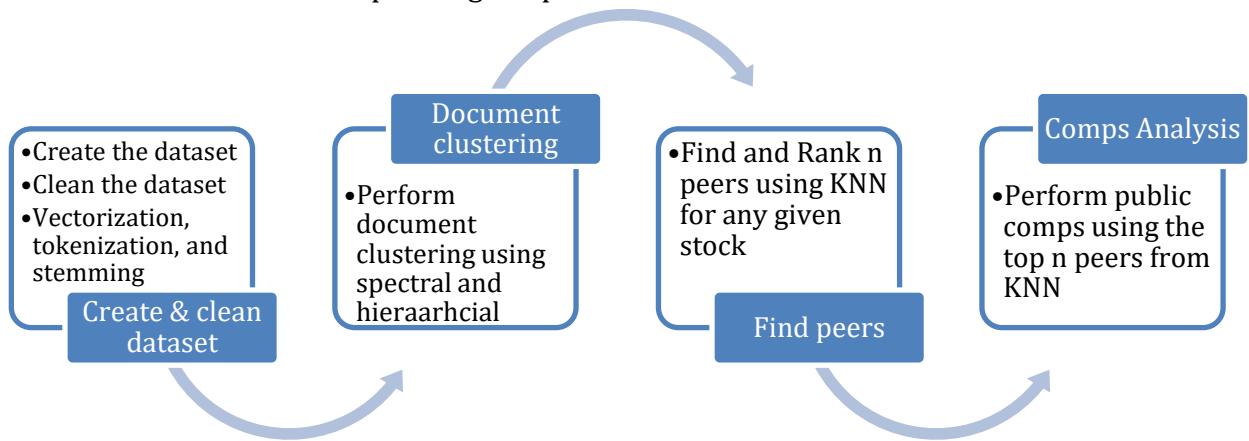

# 5. Systematic Computation of Cost of Equity for Private Corps.

Again, the idea is very simple. It is just a mere automation or extension on the traditional method to computing cost of equity for private corporations. Instead of taking the industry average as the cost of equity for a private corporation, you simply find out what its public peers are, then you take the average beta of those peers and compute cost of equity using CAPM or you can simply take their average cost of equity as the cost of equity for the private corporation. This results in more accurate cost of equity for private corporations. Let me illustrate. Suppose Microsoft was a private corporation and we wanted to compute the cost of equity for it. Microsoft operates in the "Information Technology" sector under the "Systems Software" industry. Now, if we simply took the Systems Software industry average as the cost of equity for Microsoft, we would get a value of **6.45%**. However, Microsoft's true cost of equity is **7.3%.** If we used the average of it's true competitors/peers, that may not operate in the same industry as Microsoft, such as Apple, we would get a value of **7.35%**, which is closer to Microsoft's true cost of equity compared to simply taking the industry average.

*Table 1: Microsoft's Cost of Equity from its peers generated by the Spectral model*

| Company | Cost of Equity |
|---|---:|
| Akamai | 6.30% |
| Apple | 7.55% |
| Citrix | 7.10% |
| F5 | 6.55% |
| IBM | 8.40% |
| Oracle | 8.40% |
| Salesforce | 7.15% |
| **Average** | **7.35%** |

| Microsoft | Range | Value |
|---|---|---|
| Cost of equity | 6.1% - 8.5% | 7.3% |
| Tax rate | 14.4% - 15.5% | 14.95% |
| Cost of debt | 4.0% - 4.5% | 4.25% |
| WACC | 6.0% - 8.4% | 7.2% |

*Source: https://valueinvesting.io/MSFT/valuation/wacc*

| System Software | Value |
|---|---|
| Cost of equity | 6.35% |

| System Software | Value |
|---|---|
| Tax rate | 3.36% |
| Cost of debt | 3.58% |
| WACC | 6.15% |

*Source: https://people.stern.nyu.edu/adamodar/New_Home_Page/datafile/wacc.htm*

| MSFT Peer Average | System Software Industry Average | MSFT True Cost of Equity |
|---|---|---|
| 7.35% | 6.35% | 7.3% |

# 6. Implementation
## 5.1 Creating and Cleaning the Dataset

I will be using the 487 unique stocks listed on the S&P 500 index. Some stocks have multiple classes, for example, Google has GOOG and GOOGL, hence only 487. Below is a sector distribution of the stocks listed on the index.

*Figure 1: Sector distribution of the S&P500 dataset*

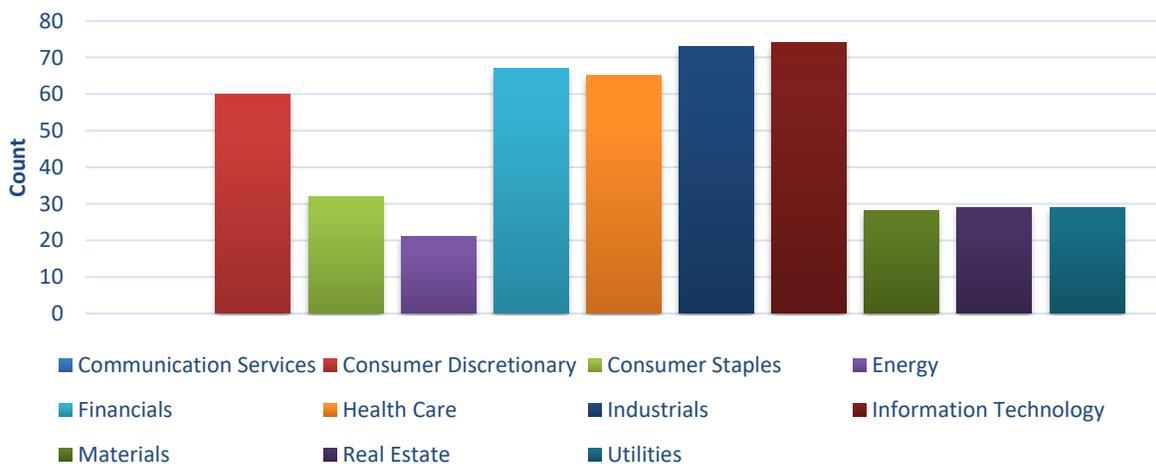

I will be using the Wikipedia pages for each of the stocks to perform document clustering. The dataset is created by scraping the Wikipedia pages. The scraped text is cleaned, removing all punctuations and non-alpha characters using NLTK. Then, the documents/pages are tokenized using NLTK tokenizer. Afterwards, all stop words are removed using the NLTK English stop words set. Next, unwanted words that do not impact the clustering such as company names like "Morgan" or "Corporation" are removed using regular expressions and manually. Finally, the cleaned tokens were lemmatized using the SnowballStemmer, which is also known as the Porter2 stemming algorithm from NLTK.

The documents are vectorized using TF-IDF Vectorizer from Sklearn (scikit learn) as it has been shown to be highly accurate and robust in performance, whilst minimizing compute power for the task of document vectorization.

### 5.2 Document Clustering using Spectral and Agglomerative Clustering

I used both Spectral and Agglomerative clustering from the Scikit learn library in Python as the clustering algorithms. For each of the model, I experimented with all combinations of the hyperparameters, including the affinity functions and different linkage types.

### 5.3 Optimizing the Models: Choosing the Optimal Number of Clusters

The models were optimized using both internal and external evaluation measures. The Silhouette coefficient was used for internal evaluation, to choose the optimal number of clusters.

Silhouette analysis can be used to study the separation distance between the resulting clusters. The silhouette plot displays a measure of how close each point in one cluster is to points in the neighboring clusters and thus provides a way to assess parameters like number of clusters visually. This measure has a range of [-1, 1]. [Sklearn]

Silhouette coefficients (as these values are referred to as) near +1 indicate that the sample is far away from the neighboring clusters. A value of 0 indicates that the sample is on or very close to the decision boundary between two neighboring clusters and negative values indicate that those samples might have been assigned to the wrong cluster. [8 Sklearn] The silhouette coefficient can be computed as follow:

## Silhouette coefficient

- *Cohesion $a(x)$*: average distance of $x$ to all other vectors in the same cluster.
- *Separation $b(x)$*: average distance of $x$ to the vectors in other clusters. Find the minimum among the clusters.
- *silhouette $s(x)$*:

$$s(x) = \frac{b(x) - a(x)}{\max\{a(x), b(x)\}}$$

- $s(x) = [-1, +1]$: -1=bad, 0=indifferent, 1=good
- Silhouette coefficient (SC):

$$SC = \frac{1}{N} \sum_{i=1}^{N} s(x)$$

Below, the silhouette analysis is used to choose an optimal value for n_clusters. The silhouette plot shows that the n_clusters value on the x-axis and the silhouette score on the y-axis. The optimal number of clusters (n_clusters) for each of the model under different hyperparameters is when the silhouette score is the highest.

*Table 2: Silhouette scores for each of the models under different parameters*

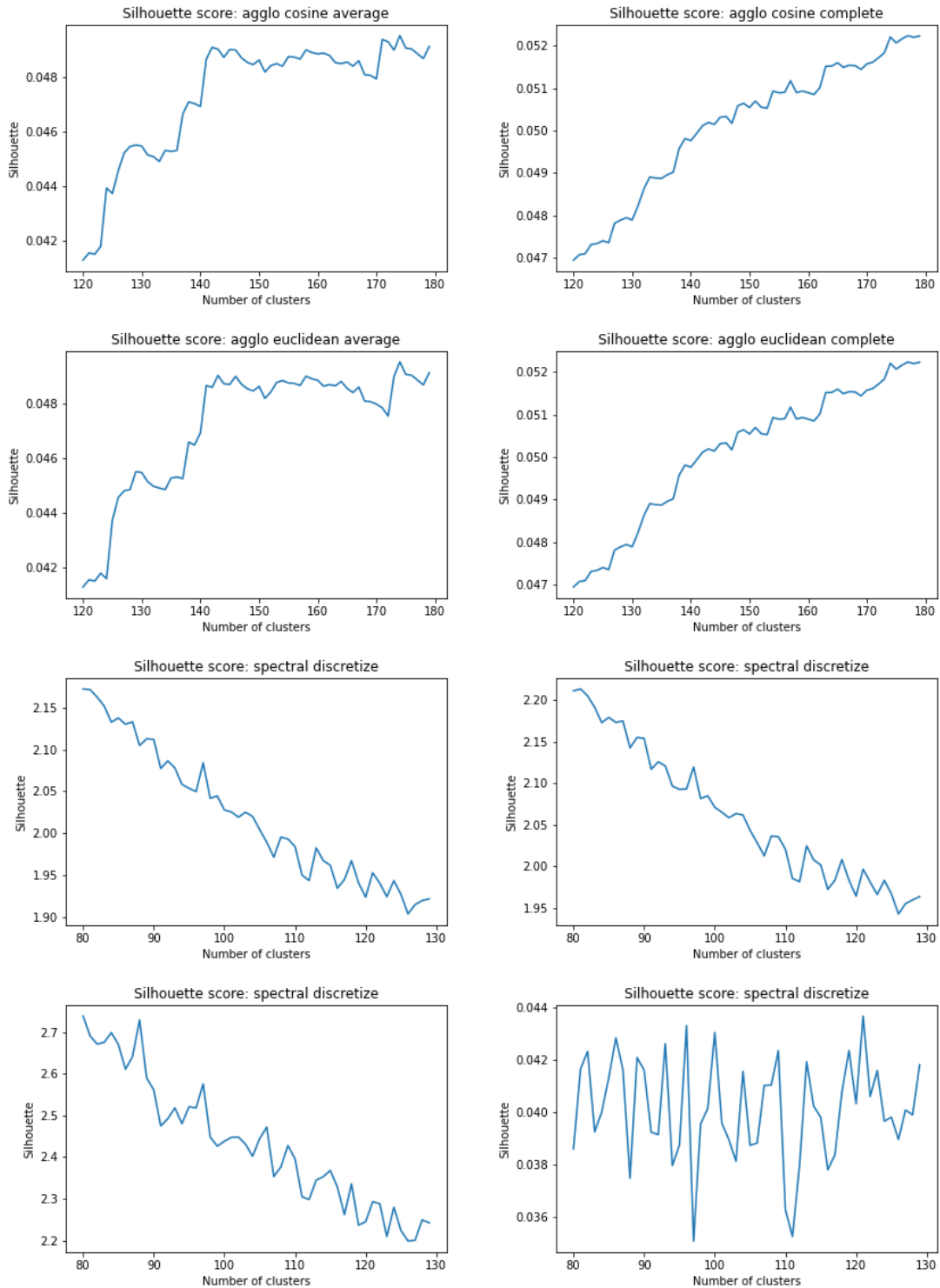

## 5.4 Finding Financial Peers using KNN

The penultimate step is finding the financial peers for the security we want to perform comps for. To accomplish this, I will be using the K-nearest neighbor algorithm. The idea is to find which of the companies in the same cluster as the company of intertest that are closest in financial measures such as Price to Earnings (P/E) ratio, Earnings Before Taxes (EBIT), and so on.

## 7. Experiments and Results

I experimented with both Spectral and Agglomerative clustering, using a number of different parameters for each of the model. For ideal clustering results, we want a slightly skewed but mostly uniform distribution for optimal clusters. The rationale behind this is, ideally, we want each company to have three to five peers at a minimum. This idealism is obviously not possible for a small dataset such as the S&P 500 index that I am using. We saw in *Figure 1 – Sector Distribution of the S&P 500 dataset*, there are a lot more companies from the Information Technology sector than the Energy Sector. From the results below, Spectral Clustering with "Discrete" label assignment results in the "best" clustering, as best is defined by a notion of idealism I described earlier; slightly skewed but mostly uniform distribution. Agglomerative clustering using the "Average" linkage and "Euclidean" affinity resulted in the worst clustering results.

Below are tree-maps showing the clustering results by Sector from the best model – the model with the highest silhouette score – which was Spectral Clustering using Discrete label assignment. At the end, there is also a tree-map showing the distribution of the clusters by Sector. Finally, further down, there are distribution plots for each of the different model with different parameters.

I performed external evaluation using a case study on Microsoft. I only performed it for Microsoft, opposed to an entire sector or the entire dataset because it would require a great deal of time to find human expert benchmark peers for every company and perform the comparisons. From the Microsoft case study in *Table 4* below, the spectral clustering model performed pretty well. There were false positives but no true negatives, precisely what we want from our models. The spectral clustering model resulted in the best results, when the silhouette score was used for optimization; choosing the best number for "n_clsuters".

*Figure 2: Spectral Clustering Results for the Communications Services Sector*

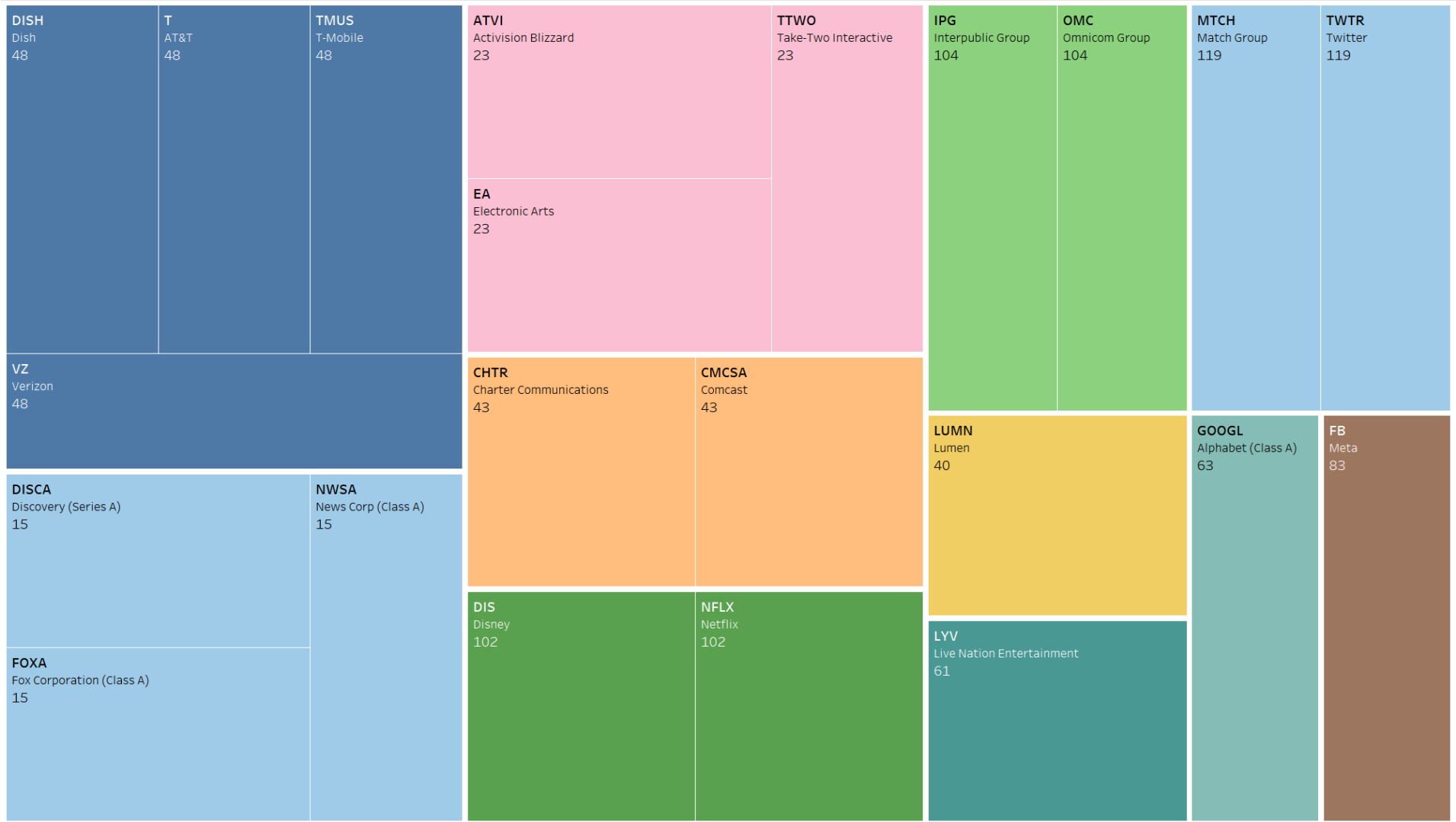

**Note:** The number is the cluster number and the color correspond to the cluster number.

*Figure 3: Spectral Clustering Results for the Consumer Discretionary Sector*

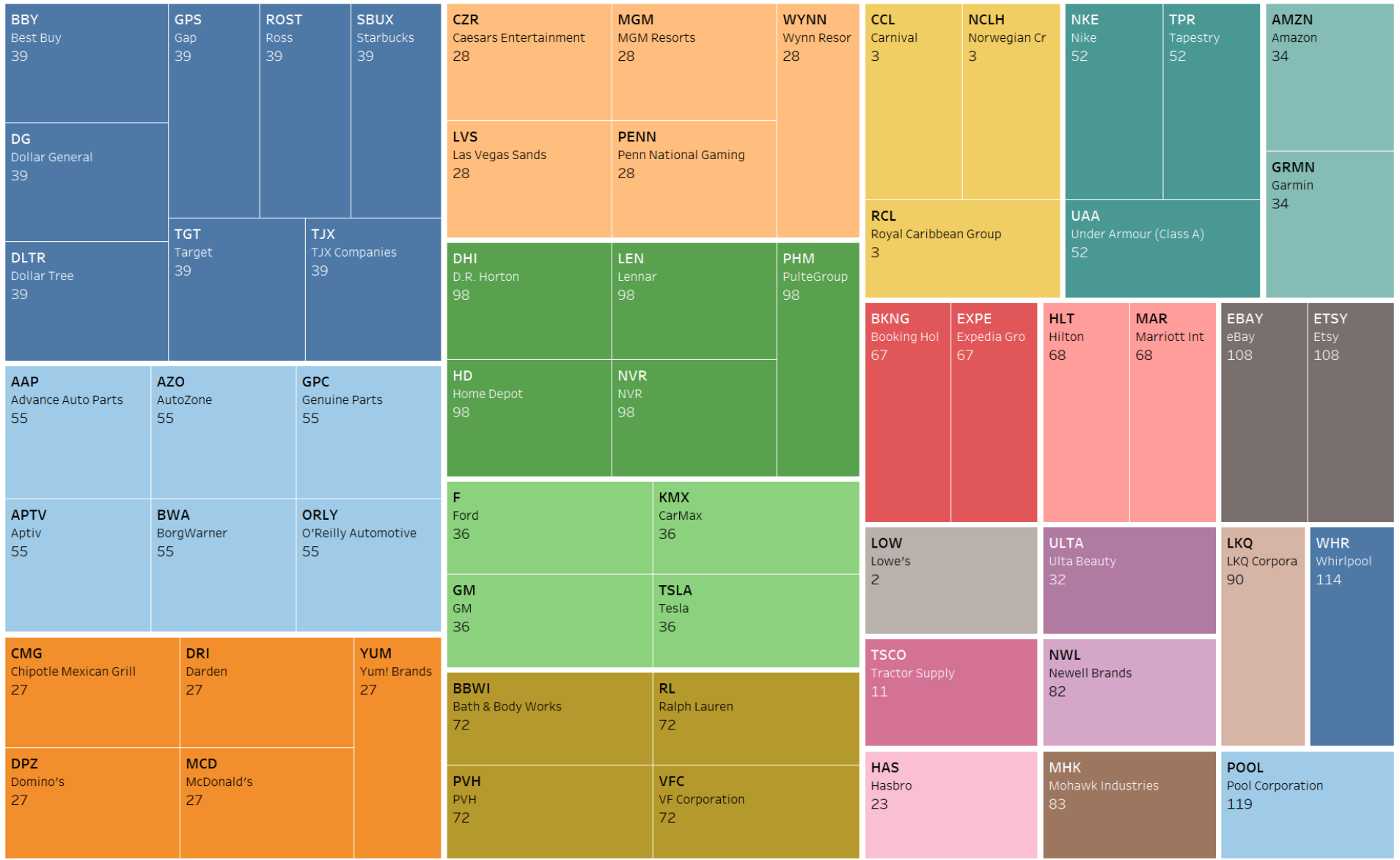

*Figure 4: Spectral Clustering Results for the Consumer Staples Sector*

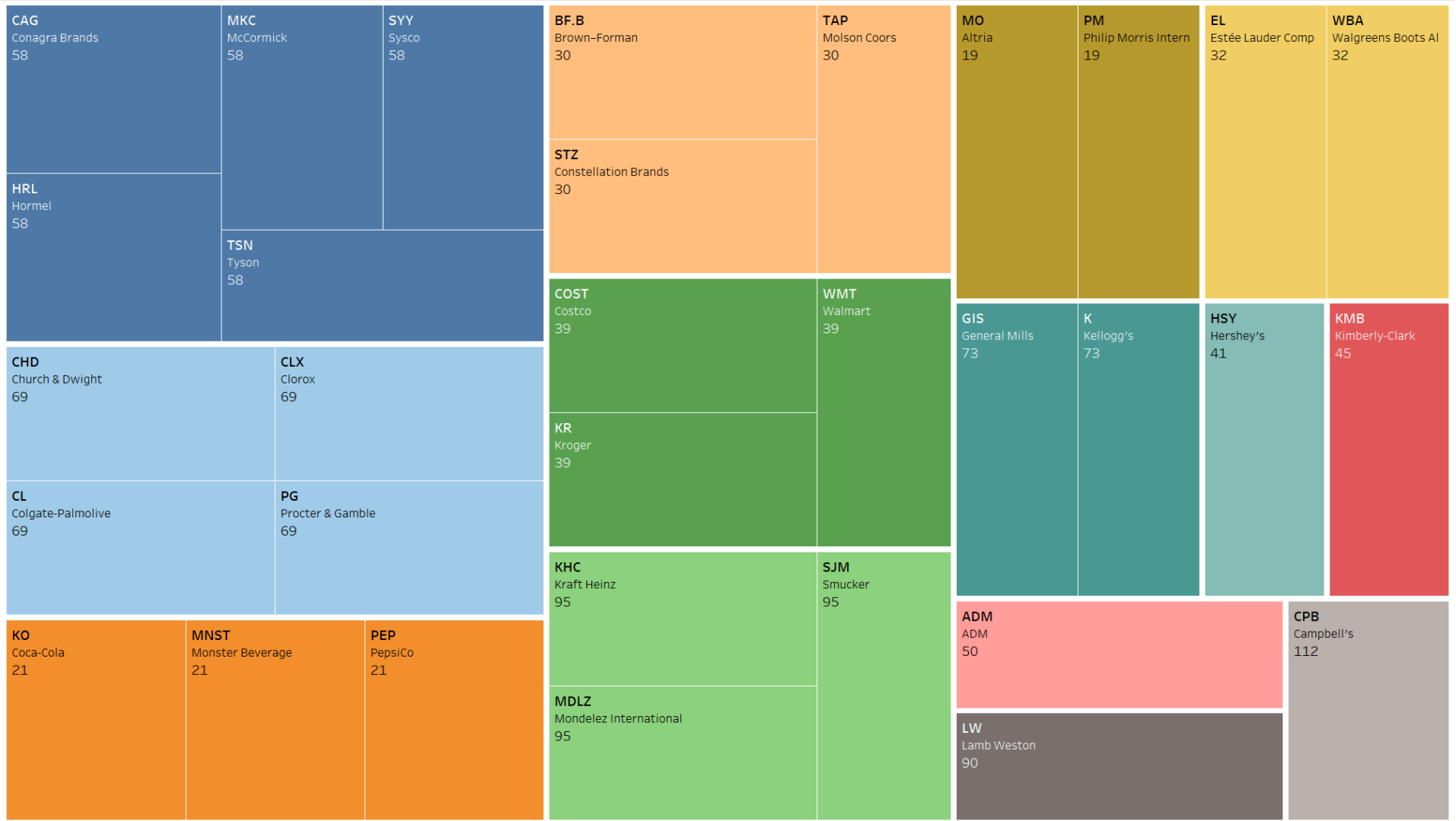

*Figure 5: Spectral Clustering Results for the Energy Sector*

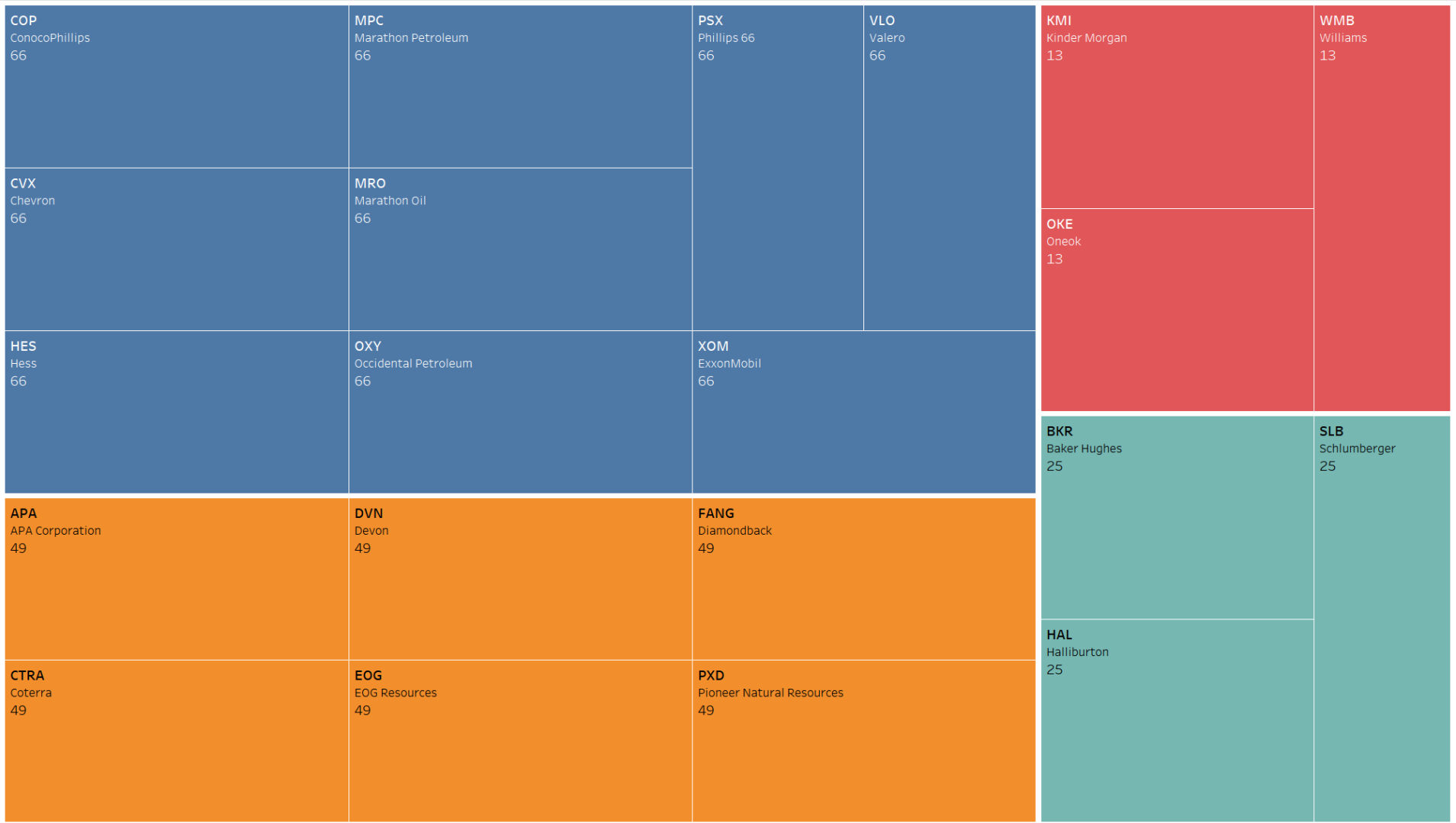

*Figure 6: Spectral Clustering Results for the Financials Sector*

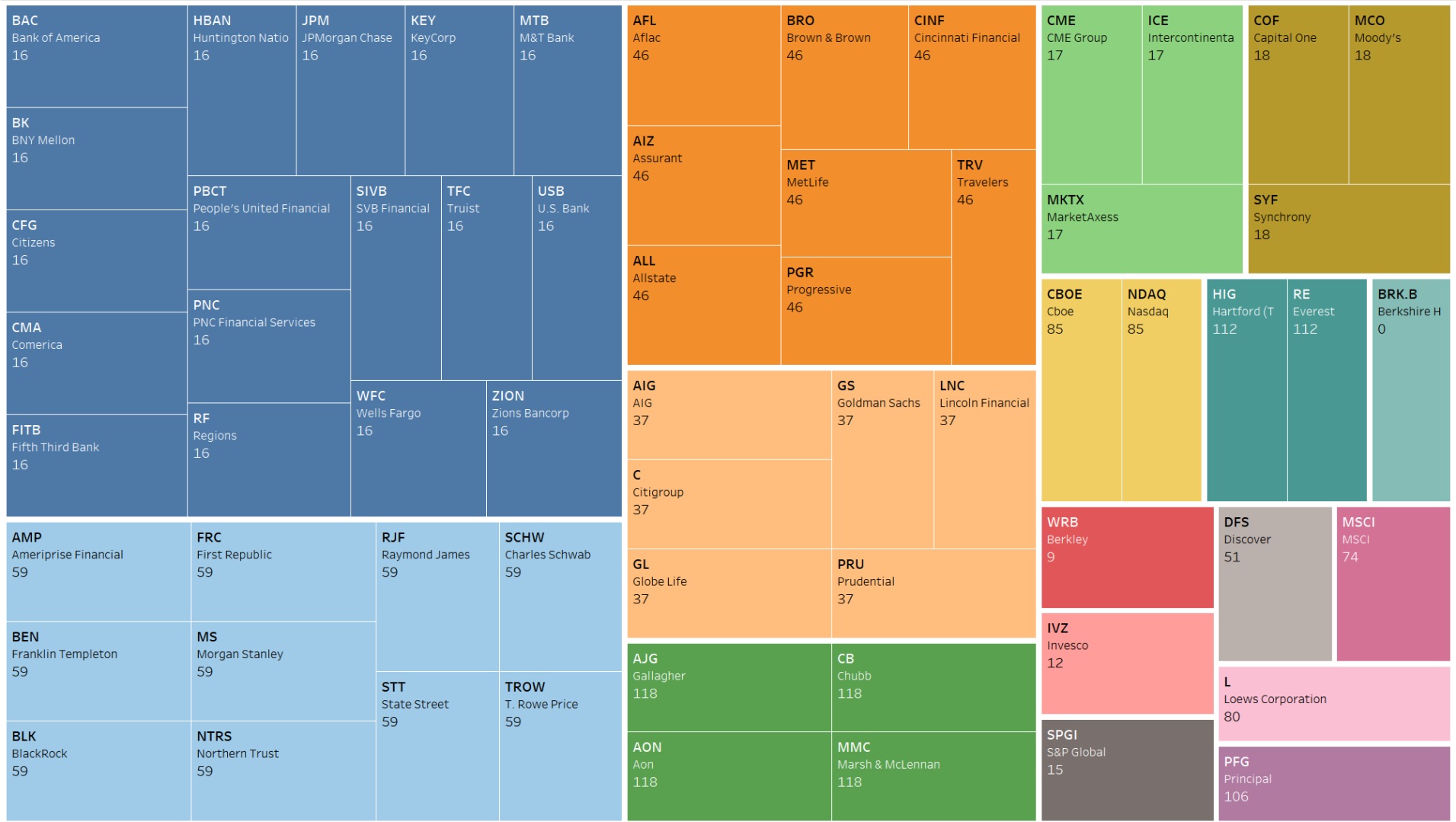

*Figure 7: Spectral Clustering Results for the Health Care Sector*

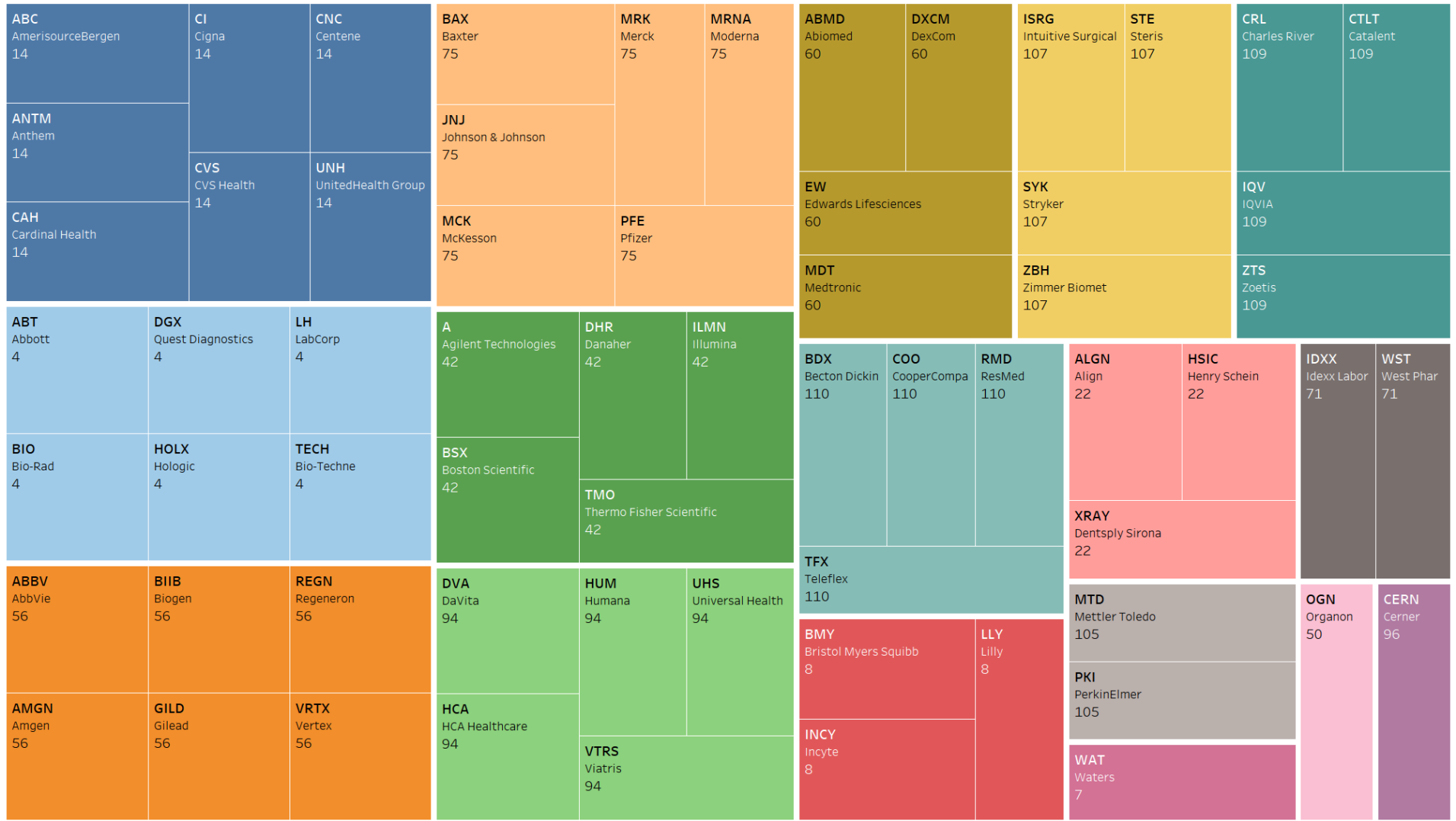

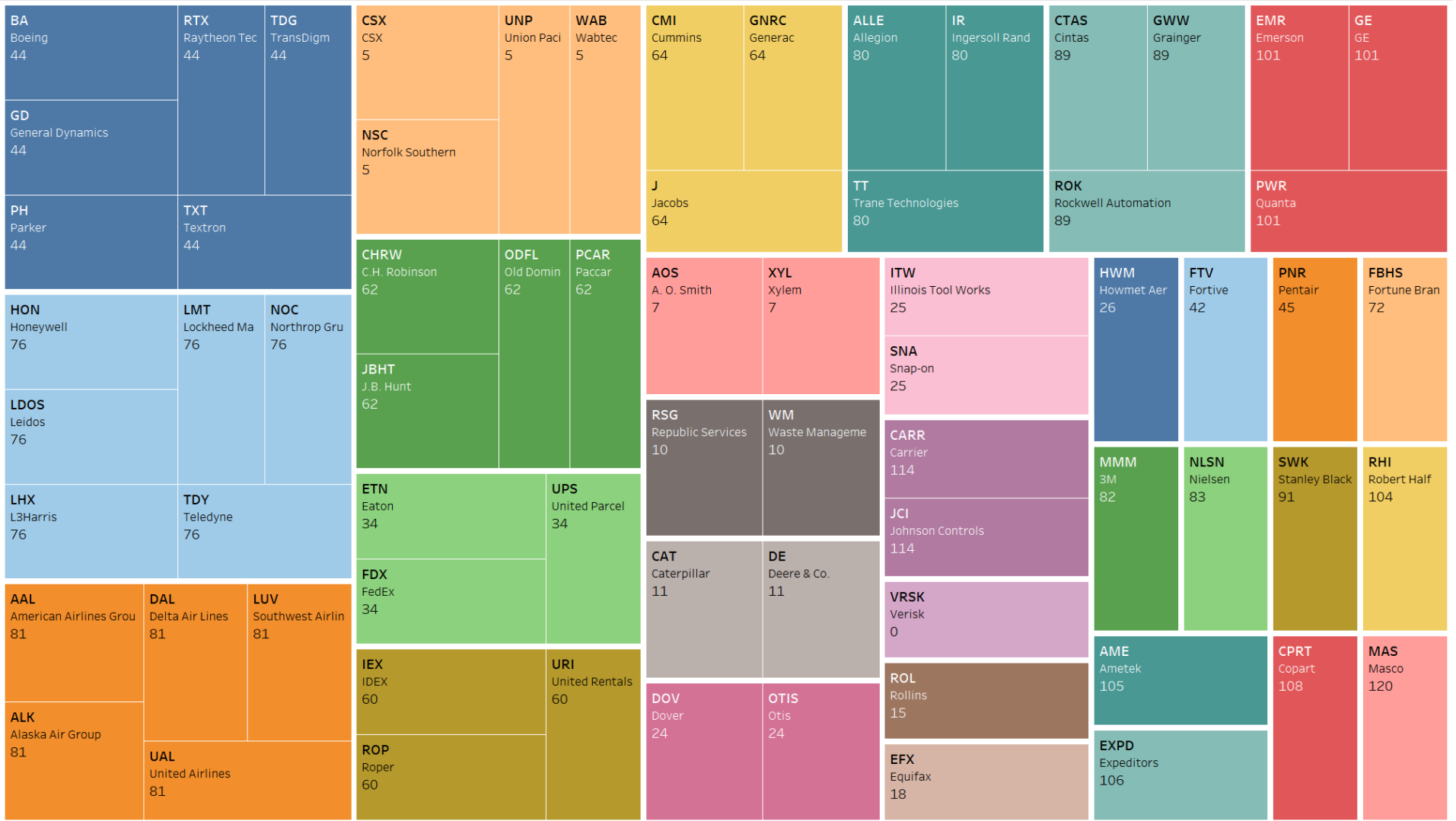

Figure 8: Spectral Clustering Results for the Industrials Sector

*Figure 9: Spectral Clustering Results for the Information Technology Sector*

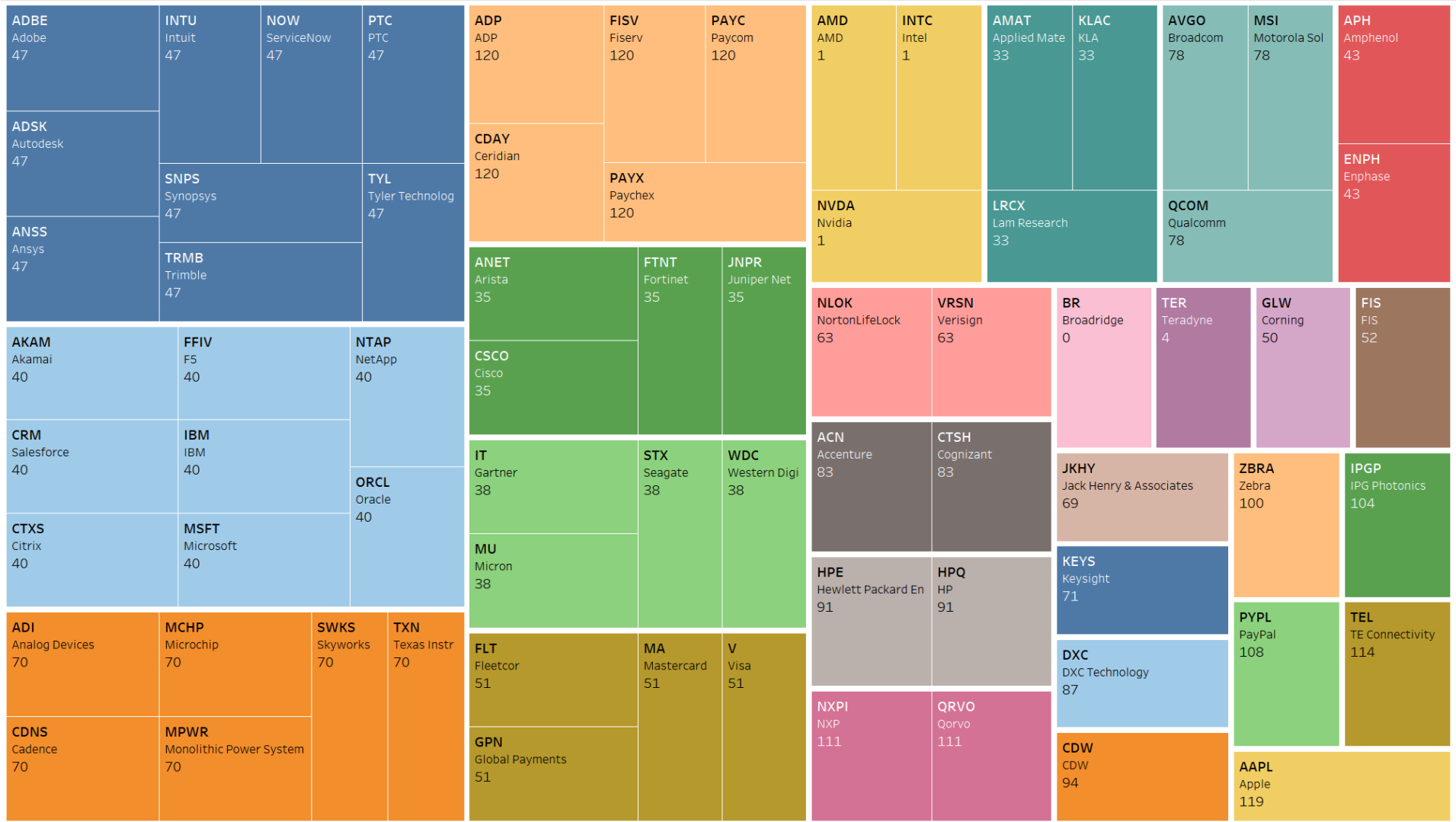

*Figure 10: Spectral Clustering Results for the Materials Sector*

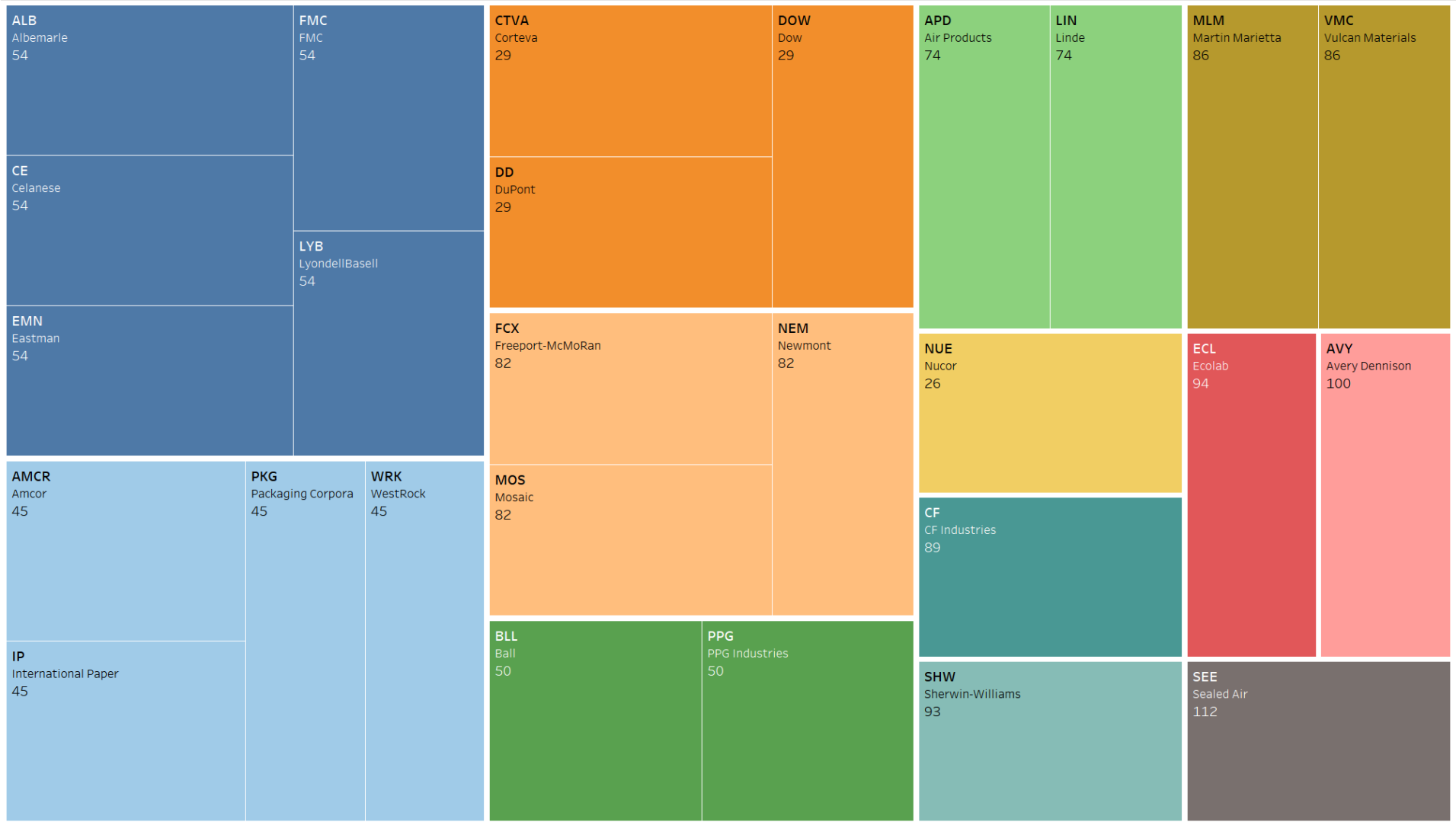

*Figure 11: Spectral Clustering Results for the Real Estate Sector*

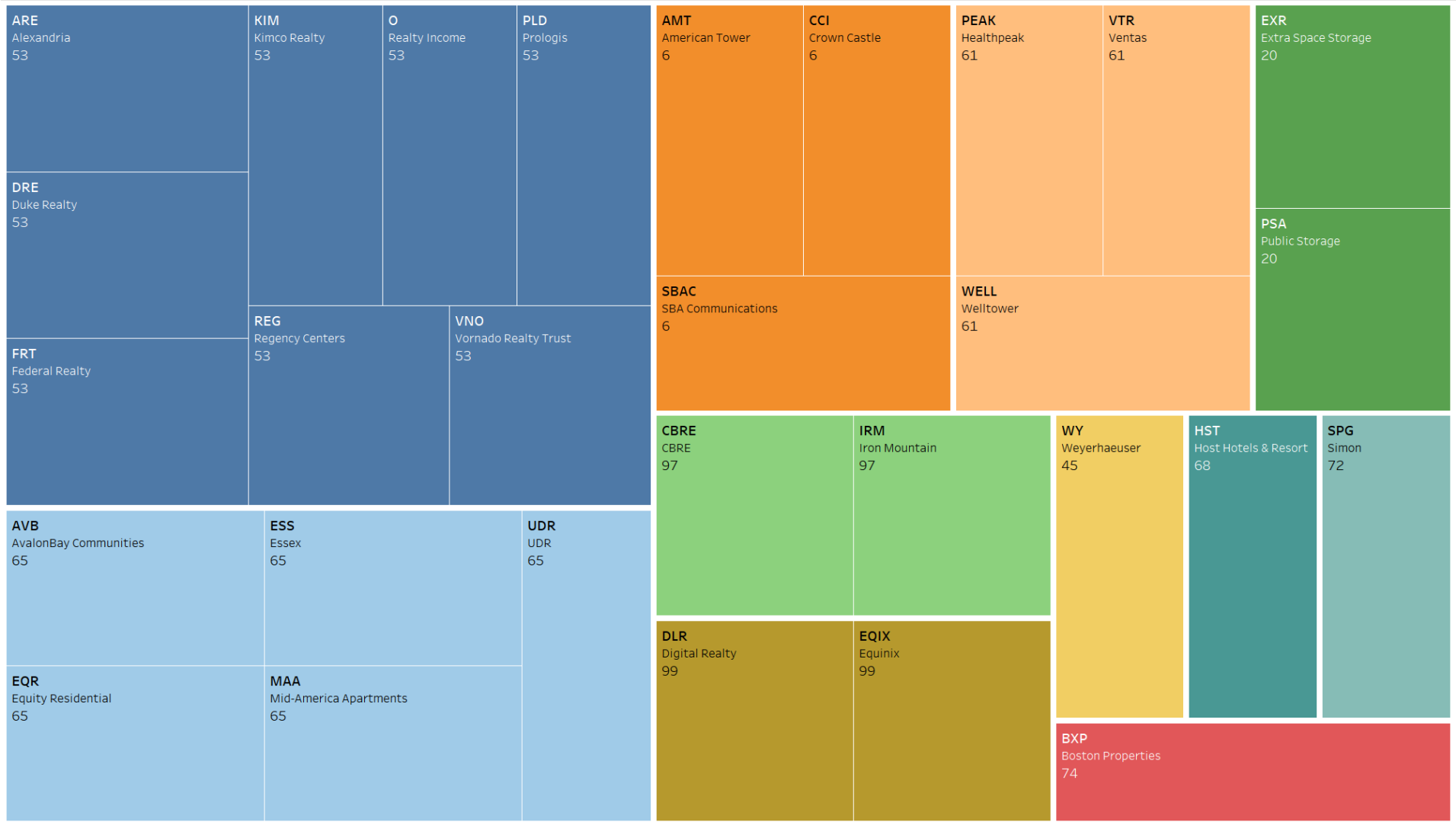

*Figure 12: Spectral Clustering Results for the Utilities Sector*

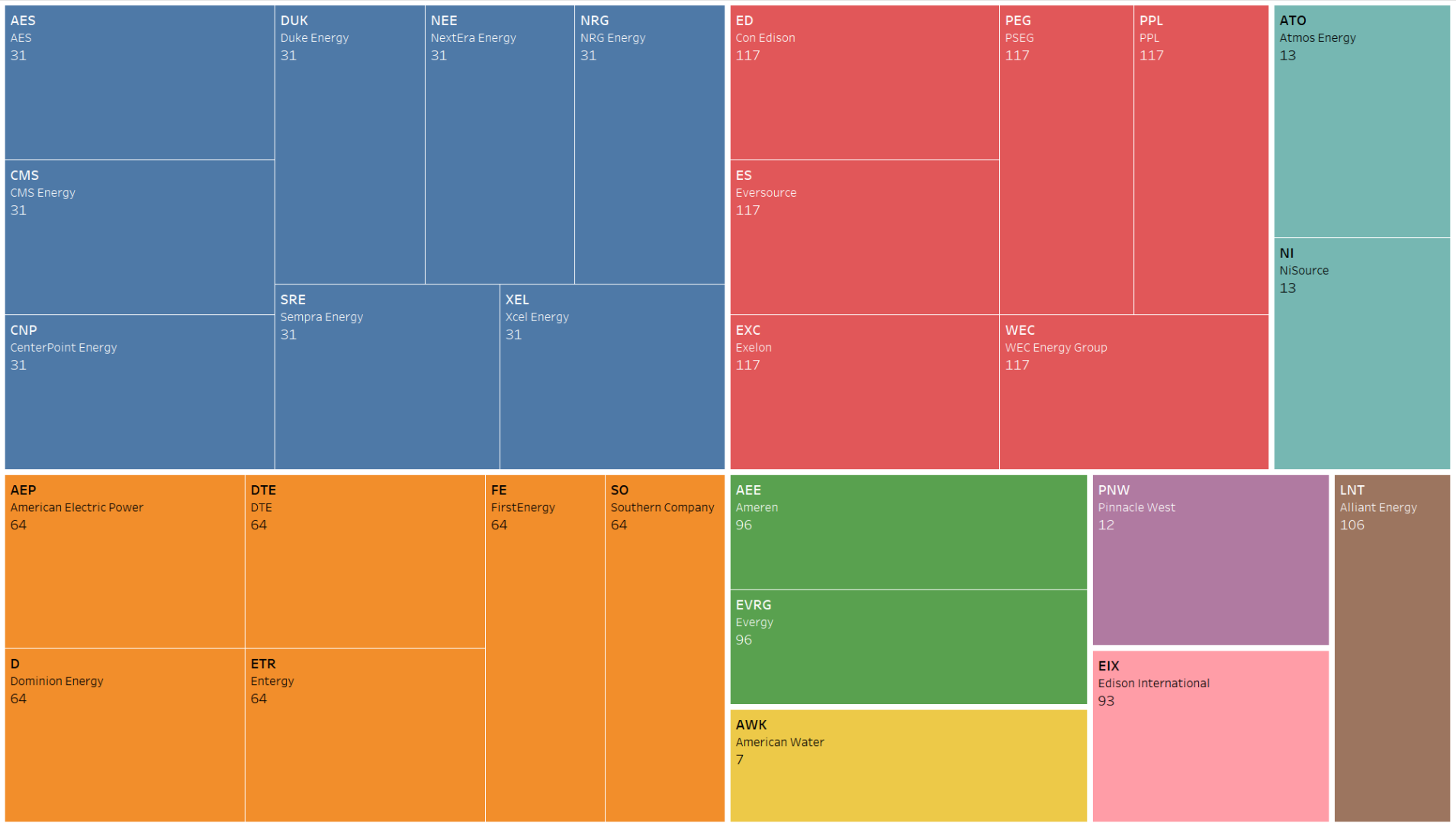

*Figure 13: Spectral Clustering Results for all the Sectors*

**Note:** The number is the cluster number and the color density represent the number of stocks under each cluster (count).

*Table 3: Cluster distributions for each of the models under different parameters*

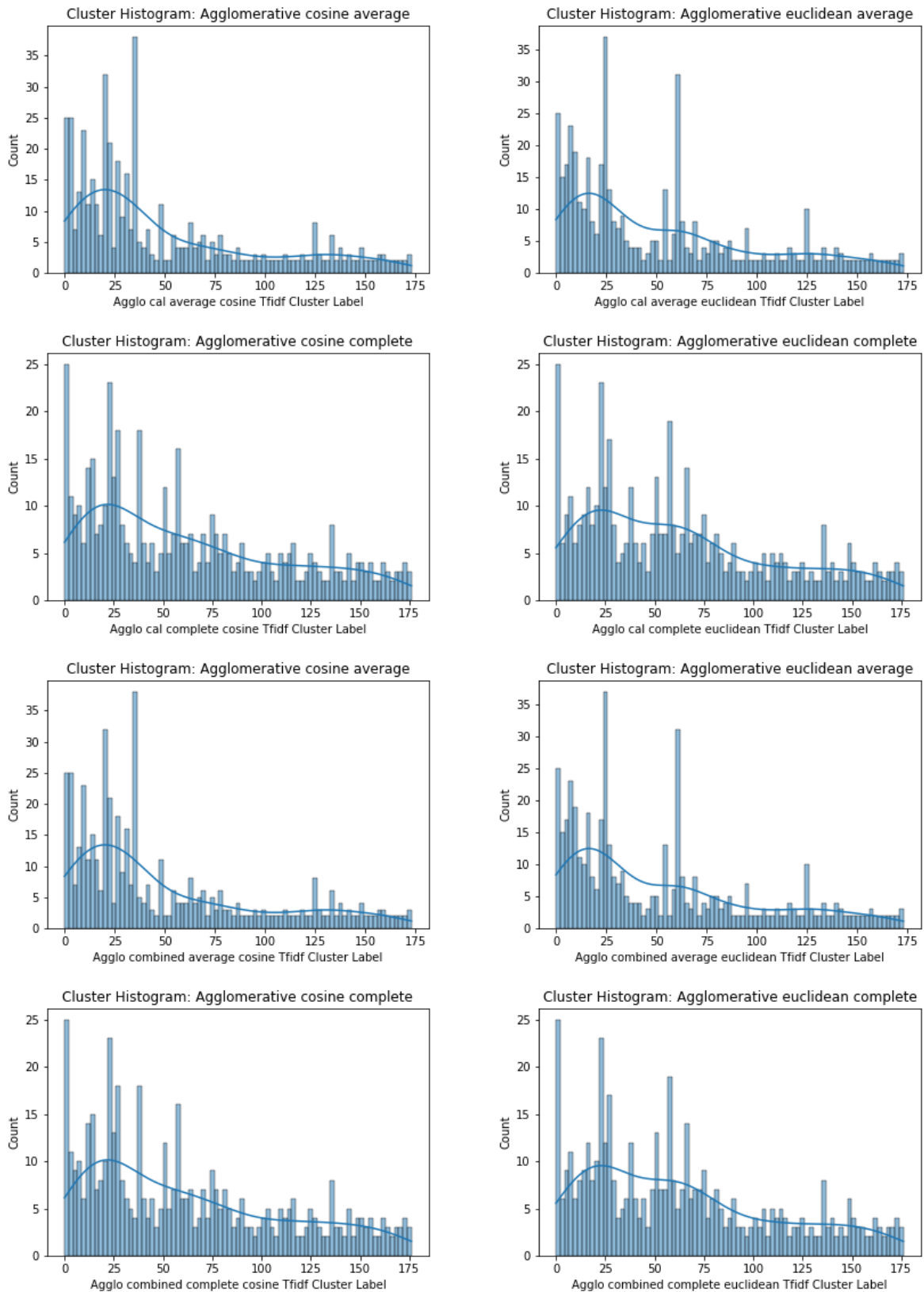

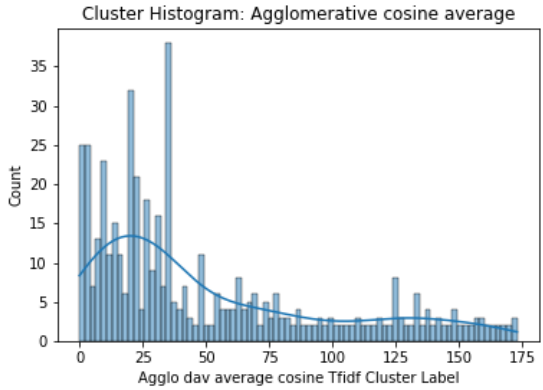
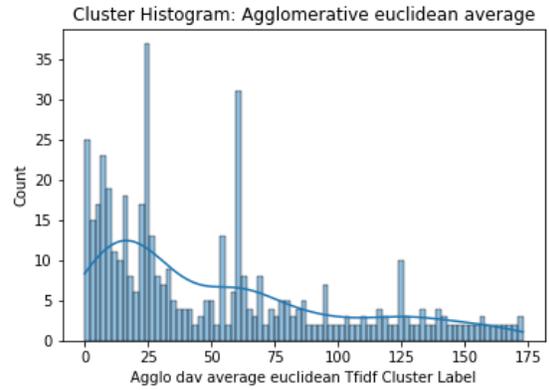
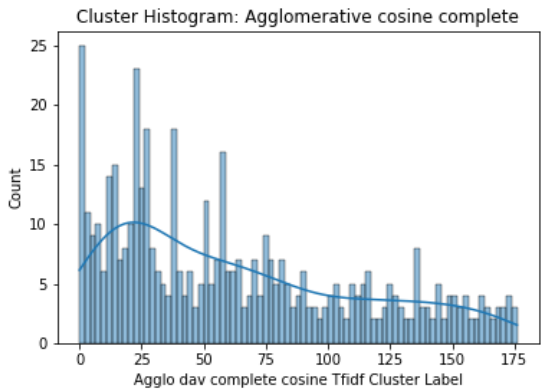
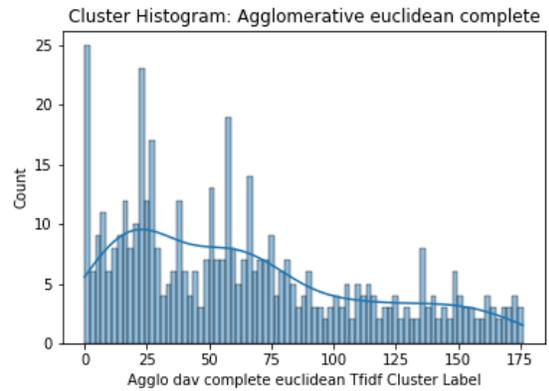
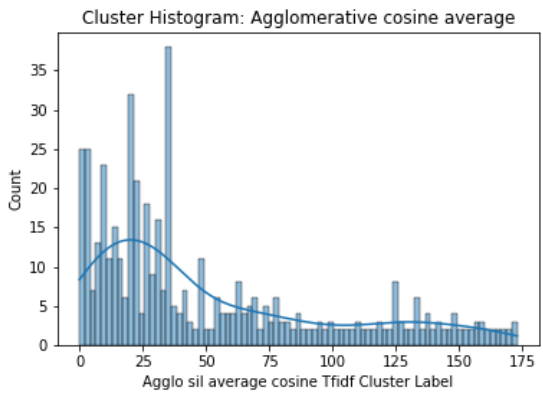
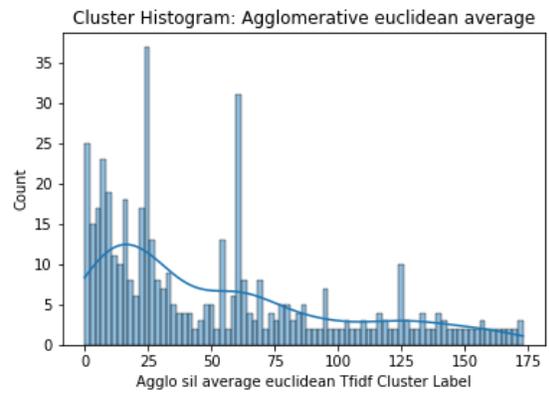
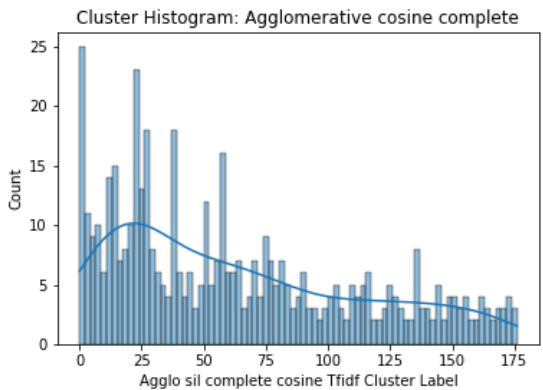
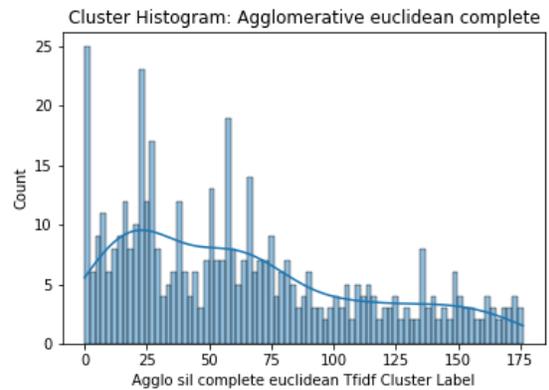

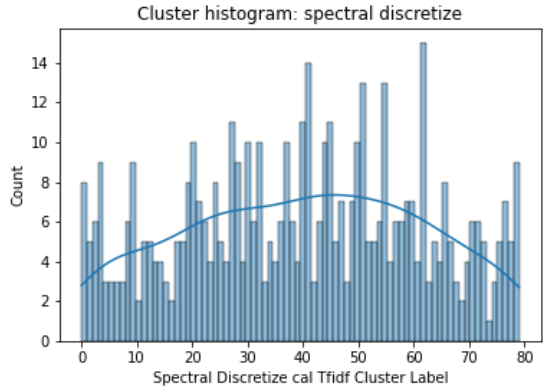
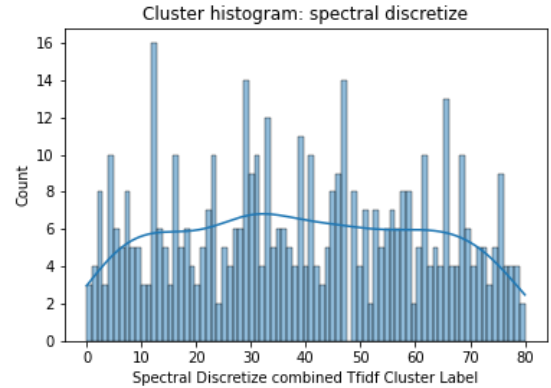
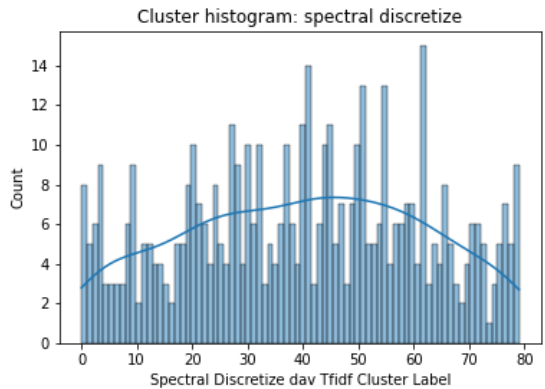
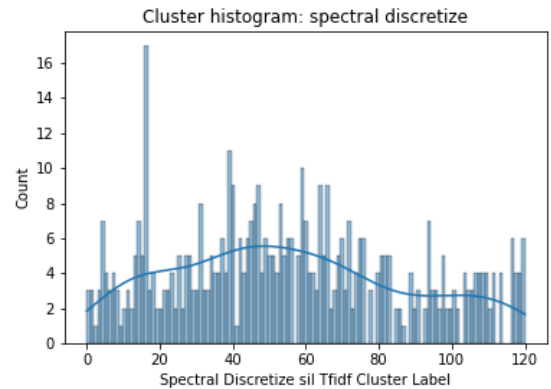

### 7.1 External Evaluations Against Human Expert Benchmark

*Table 4: Comparison of Model peers against Expert Human Peers for Microsoft*

| Spectral Clustering Peers for Microsoft | | | | | |
|---|---|---|---|---|---|
| **Silhouette** | **Calinski** | **Davies** | **Combined (Sil + Cal)** | **CapIQ** | **InvestOpia** |
| Akamai | Akamai | Akamai | Akamai | Alphabet | Apple |
| Citrix | Alphabet | Alphabet | Alphabet | Amazon | Alphabet |
| F5 | Apple | Apple | Apple | Apple | SAP * |
| IBM | Citrix | Citrix | Citrix | Oracle | IBM |
| Lumen | IBM | IBM | IBM | Salesforce | Oracle |
| Microsoft | Microsoft | Microsoft | Meta | ServiceNow | |
| NetApp | Oracle | Oracle | Microsoft | Snowflake * | |
| Oracle | Salesforce | Salesforce | Oracle | Splunk * | |
| Salesforce | Twitter | Twitter | Salesforce | VMware * | |
| | | | Twitter | Workday * | |

*To be excluded since these stocks are not in the S&P 500 dataset.

## 8. Conclusions and Future Work

To conclude, I think clustering models can successfully identify peers for companies, which can be used to systematically create comparable company analysis for both public and private companies, reducing the time required to search for peers manually. In the future, I want to explore fuzzy/soft clustering, where points can belong to more than 1 cluster, as well as deep learning models on much larger datasets using official company SEC filings. Additionally, I also want to explore differing architecture such as combing the financial and business description dataset into one.

# 9. Glossary

**Risk-Free Rate of Return**
The return expected from a risk-free investment (if computing the expected return for a US company, the 10-year Treasury note could be used).

**Beta**
The measure of systematic risk (the volatility) of the asset relative to the market. Beta can be found online or calculated by using regression: dividing the covariance of the asset and market's returns by the variance of the market.
$\beta_i < 1$: Asset i is less volatile (relative to the market)
$\beta_i = 1$: Asset i's volatility is the same rate as the market
$\beta_i > 1$: Asset i is more volatile (relative to the market)

**Expected Market Return**
This value is typically the average return of the market (which the underlying security is a part of) over a specified period of time (five to ten years is an appropriate range).